\documentclass[prb,reprint,superscriptaddress,floatfix,showpacs]{revtex4-1}
\usepackage{graphicx}
\usepackage{dcolumn}
\usepackage{bm}
\usepackage[utf8]{inputenc}
\usepackage{color}
\usepackage{hyperref}
\usepackage{siunitx}
\usepackage{amsmath}
\usepackage{ulem}
\usepackage{float}
\definecolor{purple}{RGB}{128,0,128}
\definecolor{orange}{RGB}{255,127,0}

\begin{document}
	
\title{Experimental evidence of the topological obstruction in twisted bilayer graphene. }

\author{F. Mesple}
\affiliation{Department of Physics, University of Washington, Seattle, Washington, 98195, USA}

\author{P. Mallet}
\email{email: pierre.mallet@neel.cnrs.fr}
\affiliation{CNRS, Université Grenoble Alpes, Grenoble INP-UGA, Institut Néel, Grenoble, 38000 France}

\author{G. Trambly de Laissardière}
\affiliation{Laboratoire de Physique Théorique et Modélisation (UMR 8089), CY Cergy Paris Université, CNRS, 95302 Cergy-Pontoise, France}

\author{C. Dutreix}
\affiliation{Univ. Bordeaux, CNRS, LOMA, UMR 5798, F-33400 Talence, France}

\author{G. Lapertot}
\affiliation{Univ. Grenoble Alpes, CEA, IRIG, PHELIQS, F-38000 Grenoble, France}

\author{J-Y. Veuillen}
\affiliation{CNRS, Université Grenoble Alpes, Grenoble INP-UGA, Institut Néel, Grenoble, 38000 France}

\author{V. T. Renard}
\email{email: vincent.renard@cea.fr}
\affiliation{Univ. Grenoble Alpes, CEA, IRIG, PHELIQS, F-38000 Grenoble, France}


\date{\today}
	
	

\begin{abstract}
The rich physics of magic angle twisted bilayer graphene (TBG) results from the Coulomb interactions of electrons in flat bands of non-trivial topology\cite{Bernevig2024}. While the bands' dispersion is well characterized\cite{Utama2021,Lisi2021,Nunn2023}, accessing their topology remains an experimental challenge. Recent measurements established the local density of states (LDOS) as a topological observable\cite{Brihuega2008,Mallet2012,Dutreix2016,Dutreix2019,Zhang2020,Nuckolls2023,Guan2024, holbrook2024,calugaru2025}. Here, we use scanning tunnelling microscopy to investigate the LDOS of TBG near a defect. We observe characteristic patterns resulting from the Dirac cones having the same chirality within a moiré valley\cite{deGail2011}. At higher energies, we observe the Lifshitz transition associated with the Dirac cones mixing. Our measurements provide a full characterization of TBG's band structure, confirming the main features of the continuum model including the renormalization of the Fermi velocity\cite{LopesdosSantos2007}, the role of emergent symmetries\cite{Bistritzer2011} and the topological obstruction of the wavefunctions\cite{Zou2018,Song2019}.
\end{abstract}

\maketitle

\begin{figure*}[t]
	\includegraphics[width=\textwidth]{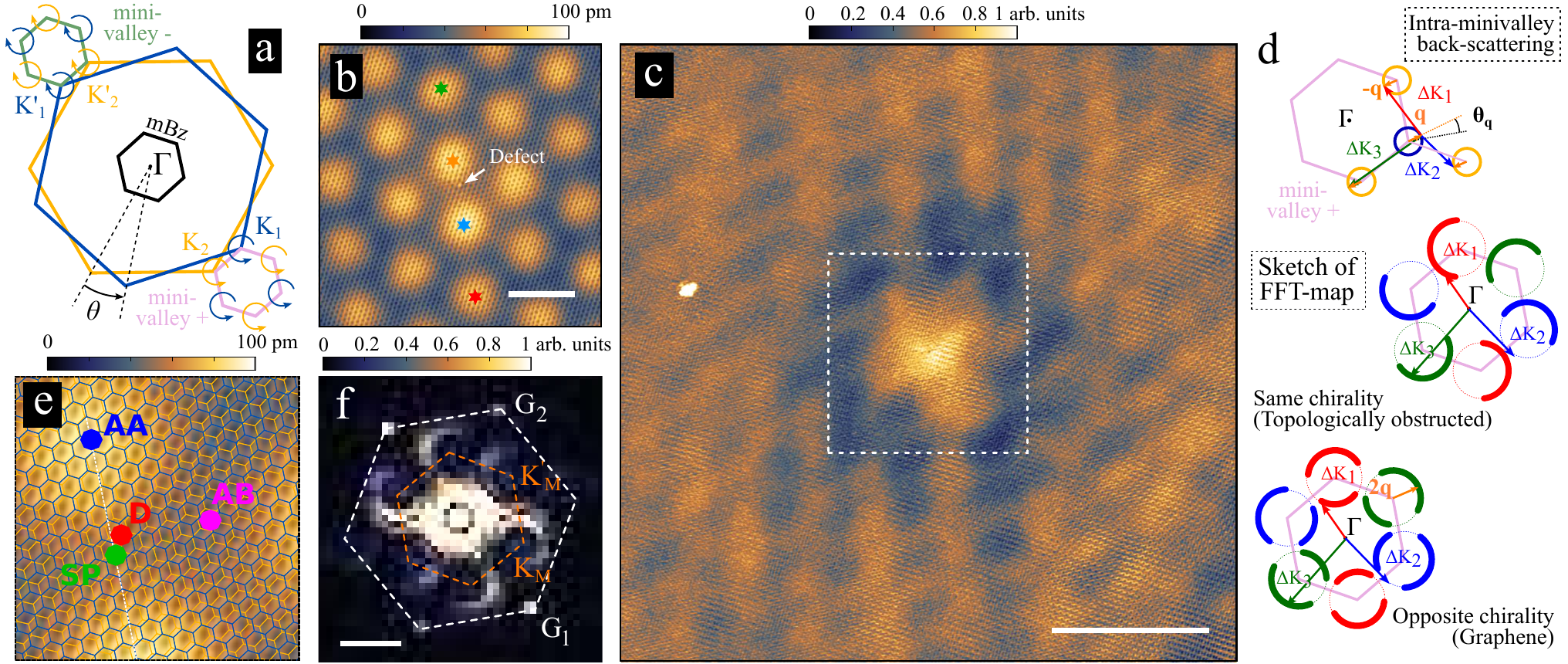}
	\caption{\textbf{Quasiparticle interference near a defect in TBG. a,} The Brillouin zones of the top (blue hexagon) and bottom layer (yellow hexagon) are rotated by the twist angle $\theta$. This defines the two moiré minivalleys (purple and green hexagons) which are far apart in reciprocal space even if they are folded on top of each other in the mini Brillouin zone (black hexagon). In each minivalley, the Dirac cones coming from different layers have the same chirality indicated by the rotating arrows. \textbf{b} STM topograph of 4.3$^\circ$ TBG ($V_b= 200$ mV, $i_t=100$ pA). A point defect is visible near the center of the image. The scale bar is 3 nm. The coloured stars correspond to the locations of the spectroscopic measurements presented in Fig.~\ref{fig: TB}a. \textbf{c,} Larger field of view of the local density of states image around the point defect at the center of the image. The image was measured at the same tunnelling conditions. The scale bar is 10 nm. The dashed square corresponds to the area of the image of panel (b). \textbf{d,} Possible intercone back-scattering within mini-valley + (top). Expected FFT of the top layer QPI signal for Dirac cones of the same (center) or opposite (bottom) chirality. \textbf{e,} The position of the observed defect is identified by the red dot and labelled \textbf{D}. The position of other defects studied in Fig.~\ref{fig: TB} is marked by dots of other colours. \textbf{f,} Modulus of the Fast Fourier Transform of the image in panel (c). The reciprocal lattice vectors $\mathbf{G_1}$ and $\mathbf{G_2}$ of the moiré are indicated and the mBz is shown in orange. The scale-bar is 1 nm$^{-1}$.}
	\label{fig: Experimental results}
\end{figure*}

The flatbands of twisted graphene layers are a fantastic playground to study the effect of strong electronic correlations leading to an extraordinary diverse gallery of phases\cite{Bernevig2024} including correlated insulators\cite{Cao2019b, Lu2019} superconductivity\cite{Cao2019,Lu2019, Stepanov2020} and strange metals\cite{Cao2020,Jaoui2022} depending on electron density, temperature, electric and magnetic fields. The flatness of the bands is not the only characteristics relevant to TBG physics. Their non-trivial topology is also crucial. This topology can be grasped in reciprocal space where the Brillouin zone of one layer is twisted with respect to that of the other (Fig.~\ref{fig: Experimental results}a) defining a mini-Brillouin zone (mBz). Since graphene has two valleys of opposite chirality, TBG also have two mini-valleys. These are independent because the moiré potential varies over length scales too large to couple them\cite{Bistritzer2011} despite they fold on top of each other in the mBz. Each mini-valley contains two Dirac cones originating from different layers and that hybridize to form the flatbands. Theory shows that the symmetry of the inter-layer hopping enforces these two Dirac cones to have the same chirality (Fig.\ref{fig: Experimental results}a)\cite{deGail2011}. This prevents the low energy description of TBG by a two Wannier orbital model which necessarily have Dirac cones of opposite chirality within a mini-valley. This is known as the topological obstruction.\cite{Zou2018,Song2019} This topology is responsible of the emergence of orbital magnet\cite{Lu2019,Serlin2020} and Chern insulator states\cite{Nuckolls2020,Wu2021,Xie2021,Yu2022} and could be involved in the superconducting state\cite{Tian2023,Tanaka2025}. It is therefore highly desirable to determine experimentally the topology of the non-interacting bands in order to settle solid foundations for the description of the strongly correlated electron physics. We have recently used the Quasi Particle Interference (QPI) pattern of the LDOS near point defects to access Dirac electrons' topology in graphene \cite{Brihuega2008,Mallet2012,Dutreix2016,Dutreix2019}. Here, following the suggestion of Ref.~\onlinecite{Mele2020}, we show that this is also relevant to determine the relative chirality of the Dirac cones within a mini-valley of TBG. 

Figure~\ref{fig: Experimental results}b shows a scanning tunneling microscope topograph of twisted graphene layers prepared at the surface of SiC as described in Ref.~\onlinecite{Kumar2016}]. The image reveals a typical moiré pattern resulting from the alternation of aligned (AA) and Bernal (AB) staking regions. The twist angle, determined from the moiré period $D$=3.25 nm, is $\theta\simeq4.3^\circ$. A defect lies at the center of the image without much influence on the rest of the image. On the contrary, it affects strongly the energy resolved LDOS image (See Methods). Figure~\ref{fig: Experimental results}c reveals periodic LDOS modulations centred on the defect, indicating that it acts as an elastic scatterer for electrons. The signal manifest in reciprocal space as circles centred on the corners $K_M$ and $K'_M$ of the mBZ (Fig.~\ref{fig: Experimental results}f). This demonstrates inter-cone scattering within a mini-valley because the wave vector of the signal links states of neighbouring Dirac cones. 
The circular shape follows from the enhanced weight of retro-diffusion in the scattering between circular contant energy contours\cite{Simon2007}. This is illustrated in Fig.~\ref{fig: Experimental results}d where coloured vectors represent the scattering between initial state $\mathbf{q}$ and final state $\mathbf{-q}$ belonging to neighbouring Dirac cones at low energies $E = \hbar v_F^* q$ ($\hbar$ is the reduced Planck constant, $v_F^*$ the Fermi velocity). Varying the orientation $\theta_{\mathbf{q}}$ of the initial state $\mathbf{q}$, the apices of these vectors trace out circles of radius $2\lVert\mathbf{q}\rVert$ around the scattering directions $\Delta\mathbf{K}_{1}$, $\Delta\mathbf{K}_{2}$, and $\Delta\mathbf{K}_{3}$ (Fig.~\ref{fig: Experimental results}d). However, we observe only part of the expected $2q$-circles which therefore appear as $2q-$arcs. We will now show that this is how the topological obstruction of wave functions manifests in this interferogram.

For twist angles below 10$^\circ$, the electronic structure of TBG supports a four-band continuum description in which the electrons behave as massless relativistic fermions with identical chirality in a given mini-valley. The moiré potential renormalises the Fermi velocity \cite{LopesdosSantos2007,Bistritzer2011,deGail2011}. This non-interacting band structure can be probed at the experimental twist angle since the bands are still dispersing\cite{Mele2020}. The wave functions characterize the charge distribution in the moir\'e unit cell and capture the four sublattice degrees of freedom. This allows for the description of scatterers localized at the scale of the moiré lattice as in our experiment. We determine the LDOS modulations in the upper layer due to  back-scattering between nearest-neighbour Dirac cones of same chirality (Supplementary information). At large distances from the scatterer, they exhibit the universal oscillations
\begin{align}\label{Eq : LDOS_Theo}
	\delta \rho (\mathbf{r},\mathbf{\Delta K}_{n}) 
	\propto 
	\mathcal{I}_{n}(\theta_{\mathbf{q}}) \,
	\text{Re}
	\left[
	\frac{e^{i(\mathbf{\Delta K}_{n}-2\mathbf{q})\cdot (\mathbf{r}-\mathbf{r}_{0})}}{q_{0}|\mathbf{r}-\mathbf{r}_{0}|}
	\mathcal{T}
	\right] \,.
\end{align}
where the complex number $\mathcal{T}$ captures the elastic scattering in all orders and depends on the specific position $\mathbf{r_{0}}$, sublattice structure, symmetry, and strength of the defect. The algebraically decaying oscillations give rise to a $2q$-circle centred on $\mathbf{\Delta K}_{n}$ in the FFT. The intensity along the $2q$-circles is modulated by $\mathcal{I}_{n}(\theta_{\mathbf{q}})=\xi + \chi \cos\left(\theta_{\mathbf{q}}+\phi_{n}\right)$. This intensity factor results from the STM tip coupling to the incoming and outgoing wave functions in the upper layer (supplementary information). The band index satisfies $\chi=+1$ ($-1$) in the conduction (valence) band and the valley index $\xi=\pm1$ denotes the chirality. The phase $\phi_{n}=(1-n)2\pi/3$ depends on the orientation of the scattering wave-vector $\pm\Delta\mathbf{K}_{n}$. Remarkably, the intensity factor $\mathcal{I}_{n}$ predicts zeros of intensity that are independent of the defect specificities. Along the $2q$-circles, $\mathcal{I}_{n}$ vanishes once at the angle $\theta_{\mathbf{q}}=\pi-\phi_{n}\,[2\pi]$. The QPI contribution from the mini-valley of opposite chirality exhibits the same behavior so that the signal of the two valleys simply add-up (Supplementary information). The QPI expected within the four-band continuum description aligns perfectly with the experiments. We note that the $2q-$arcs were not resolved in Ref.~\onlinecite{Mele2020} which considered the sum of the LDOSs of the two layers, while the STM tip only probes that of the top layer. 

\begin{figure*}[ht!] 
	\centering
	\includegraphics[width=\textwidth]{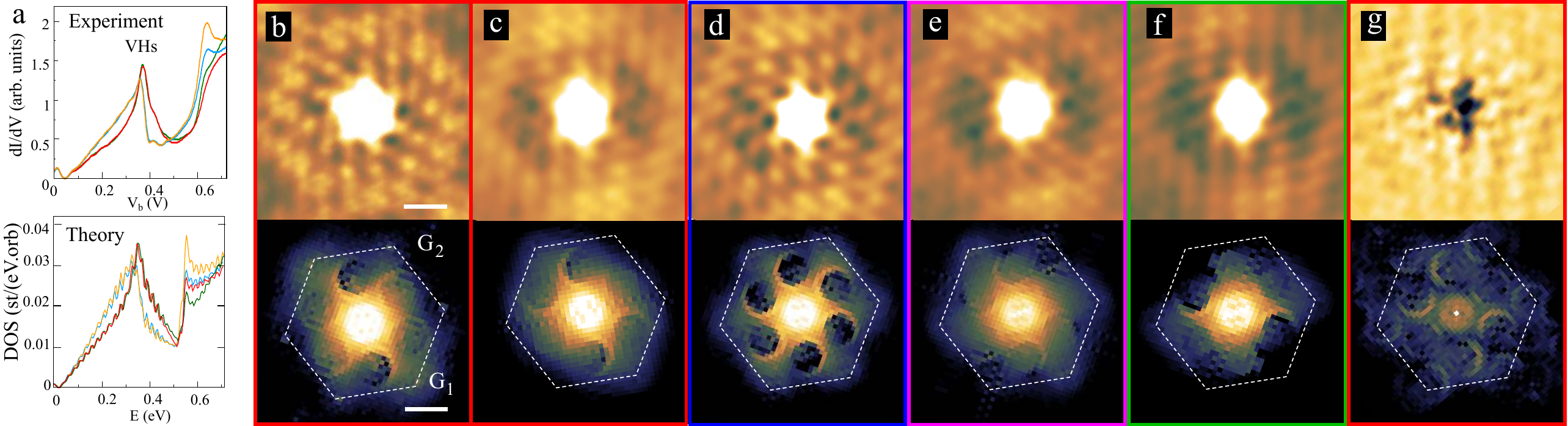}
	\caption{\textbf{Independence on the defect's nature and position}. \textbf{a,} Experimental local density of states measured from the dI/dV spectroscopy (top). Corresponding theoretical LDOS (Bottom). The different spectra were measured at the locations indicated by a star of the same color in Fig.~\ref{fig: Experimental results}b. \textbf{b, c, d, e, f, g} Tight-binding calculations of the QPI signal in the surface layer near a defect (top) and it Fourier transform (bottom). The scale bar is 10 nm in the realspace and 1 nm$^{-1}$ in reciprocal space images. In \textbf{b-f}, the defect is modelled by a Gaussian potential of depth 1.5 eV and width $\sigma=1.5 a_0$ ($a_0$ is graphene's lattice constant). In \textbf{g}, the defect is a non-reconstructed carbon vacancy located in the top layer. The defect is at position $\textbf{D}$ in panels \textbf{b,c} and \textbf{g}, AA in panel \textbf{d}, AB in panel \textbf{e}, SP in panel \textbf{f}. See Fig.~\ref{fig: Experimental results}e for the positions of these defects. In \textbf{c-g}, the model includes heterostrain. All the results are displayed for an energy 175 meV below the van Hove singularity as in the measurement of Fig.~\ref{fig: Experimental results}c.} 
	\label{fig: TB}
\end{figure*}

Alternativelly, TBG can also be modelled by exponentially localized Wannier orbitals centred at the AB and BA regions\cite{Kang2018,Koshino2018}. The corresponding two-band continuum model produces similar band dispersion but with two Dirac cones of opposite chiralities within in each mini-valley. A localized scatterer would then induce LDOS modulations with the same asymptotic behavior as in Eq.\ref{Eq : LDOS_Theo}, but with a modified intensity factor $\mathcal{I}_{n}(\theta_{\mathbf{q}})=1+\cos\left(2\theta_{\mathbf{q}}+\phi_{n}\right)$ (Supplementary information). Consequently, the FFT of the QPI patterns would also consist of $2q$-circles, but with two extinctions as was observed in graphene\cite{Brihuega2008,Mallet2012,Dutreix2021}.
Therefore, the observation of $2q$-arcs is compelling evidence that the Dirac cones within a mini-valley have the same chirality and cannot derive from a two-band Wannier representation, thus confirming the topological obstruction of the wave functions.

We perform tight-binding calculations\cite{Trambly2010,Huder2018} to investigate numerically this obstruction (see supplementary information for details). We first confirm the relative chirality of Dirac cones by computing the Berry curvature $\Omega$.\cite{Fukui2005}  We find it to be singular like in graphene and it is the same for Dirac cones within a mini-valley consistent with the continuum model (Fig. 3c). We then investigate if our experiments can be reproduced by the numerical model. The position of the defect within the moiré unit cell is determined from Fig.~\ref{fig: Experimental results}e. We combined the spatial dependence of the LDOS and calculated QPI images to determine the best description of the scatterer. Figure~\ref{fig: TB}a and \ref{fig: TB}b shows that the scatterer is faithfully modelled by a -1.5eV deep Gaussian potential extended over $\sigma= 1.5 a_0$ in the two layers (See supplementary information for a discussion of other defects). The size is compatible with experimental data (Fig.~\ref{fig: Experimental results}b) and justifies the assumptions of a point scatterer localised at the moiré scale. Furthermore, the calculations of Fig.~\ref{fig: TB}d-f, show that the $2q$-arcs are robust against the position of the defect within the moiré unit cell and against the nature of the defect (Fig.~\ref{fig: TB}g). This confirms independently the universal character of the intensity factor of the continuum model and strongly establish our observation as an experimental proof of the topological obstruction.

\begin{figure*}[t]
	\centering
	\includegraphics[width=\textwidth]{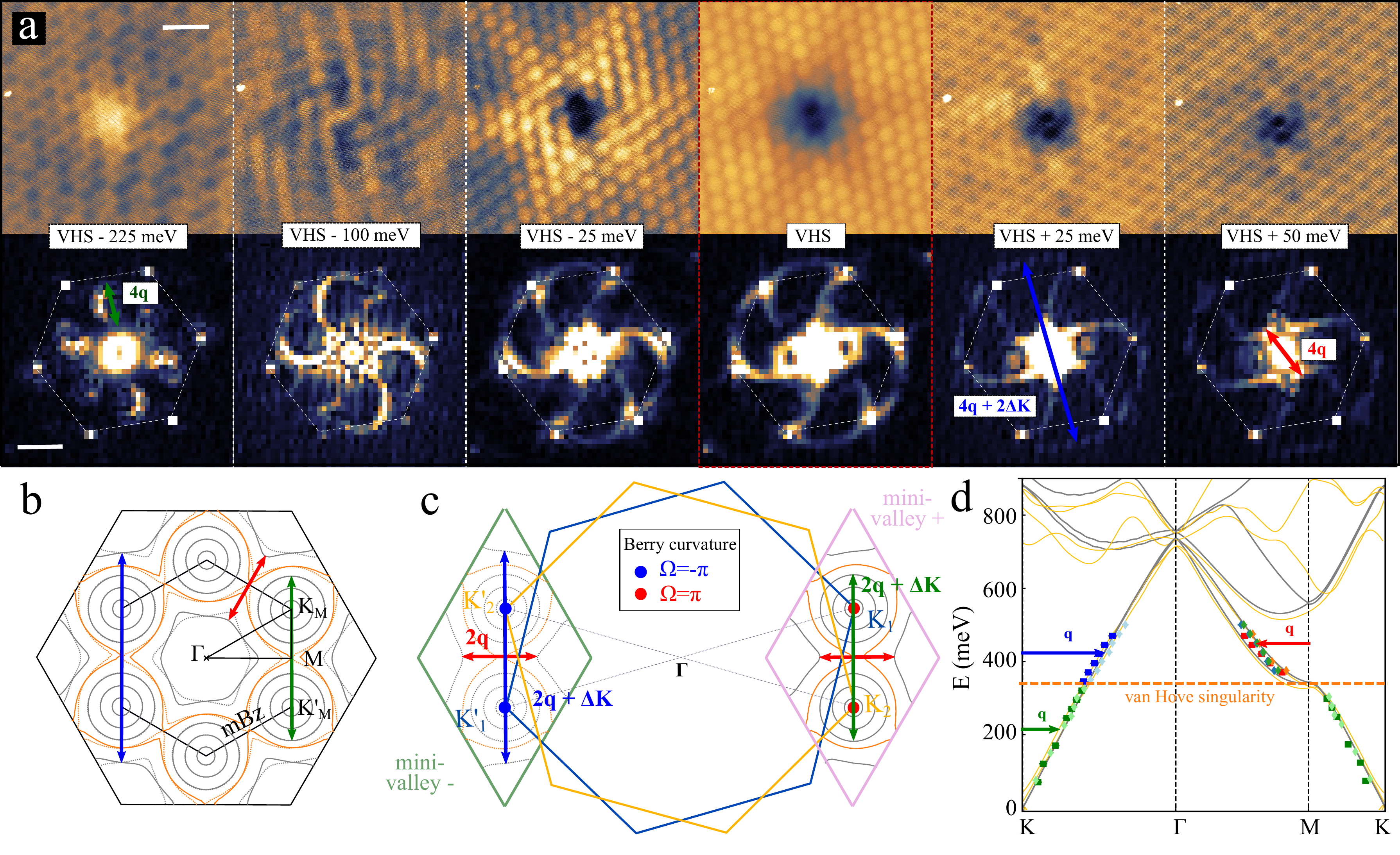}
	\caption{\textbf{Lifshitz transition at the van Hove Singularity. a,} Full energy dependence of the experimental QPI signal both in real (top row) and reciprocal space (bottom row).  The scale bars are 10 nm and 1 nm$^{-1}$ respectively. \textbf{b,} Calculated constant energy contours of the pristine (without defect) TBG lowest energy bands across the Lifshitz transition. The coloured arrows indicate the possible backscattering processes. which can be measured from the experiments of Fig.~\ref{fig: Lifshitz}a.  \textbf{c,} Same data unfolded to the original graphene valleys to explicit the intra-minivalley scattering process represented by coloured arrows. The twist angle has been increased for presentation purpose. \textbf{d,} Energy dispersion of TBG reconstructed from the data of Fig.~\ref{fig: Lifshitz}a (solid squares) and the corresponding images calculated by tight-binding (diamonds). See supplementary information for these images. These are overlaid to the electronic dispersion calculated by tight-binding with (yellow) or without (black) heterostrain. The coloured arrows correspond to the scattering processes introduced in Fig.~\ref{fig: Lifshitz}b.} 
	\label{fig: Lifshitz}
\end{figure*}

The energy dependence of QPIs displayed in Fig.~\ref{fig: Lifshitz} can be exploited to explore the entire band structure with a momentum resolution of 0.01\AA$^{-1}$ and energy resolution $\sim$10 meV. At low energies, we use the radius of the arcs to retrieve the linear dispersion as was done in graphene\cite{Mallet2012} (Fig.~\ref{fig: Lifshitz}d). We find that the Fermi velocity is renormalised to $0.95\times10^6$m/s. This is very close to the prediction of the tight-binding calculations from which we expect $v_F^* = 0.85v_{mono} = 0.91\times10^6 $m/s at this twist angle. Here, $v_{mono}$ is the Fermi velocity of monolayer graphene. At higher energies, the Dirac cones anti-cross at the $M$ point of the mBz forming a van Hove singularity (VHS) at which the Fermi surface undergoes a Lifshitz transition switching to a star shaped contour around the $\Gamma$ point (Fig.~\ref{fig: Lifshitz}b). This is reflected in Fig.~\ref{fig: Lifshitz} where a star shaped signal arises in the FFT-QPI signal which is well reproduced by the tight-binding calculations (supplementary information). From these data and the corresponding backscattering processes described by red and blue arrows in Fig.~\ref{fig: Lifshitz}a, \ref{fig: Lifshitz}b and \ref{fig: Lifshitz}c we can measure the dispersion above the van Hove singularity. We compare the experimental data to the results of the numerical model in Fig.~\ref{fig: Lifshitz}d and find an overall very good agreement. 

In order to provide the most accurate description of our experiments, we will now discuss the effect of heterostrain which is ubiquitous in TBG\cite{Huder2018,Mesple2021}. Our analysis presented in the supplementary information reveals a small uniaxial heterostrain of 0.2\% and a biaxial heterostrain of 0.4 \%. Tight-binding calculations including the full relative arrangement show that this induces a slight distortion of the low energy bands  (Fig.~\ref{fig: Lifshitz}d). However, this does not alter the extinctions in the QPI rings (Fig.~\ref{fig: TB}c-g and supplementary information) providing a hint that the topology of TBG is robust against heterostrain as recently proposed\cite{Herzog2024}. It will be interesting to verify that this is still the case at lower twist angle where strain has an increasing effect\cite{Bi2019,Mesple2021}.\\ 
More importantly, our analysis of the relative arrangement of the layers demonstrates that the moiré lattice is not commensurate with the graphene lattice  (supplementary informations). This means that we are certain that translation symmetry is only approximate at the moiré scale. Yet, the continuum model describes our experiments well despite it accounts only for emerging symmetries at the moiré scale and not the exact symmetries at the atomic scale.\cite{Bistritzer2011,LopesdosSantos2012, Zou2018}.      

Finally, we have measured the QPI signal of other point defects in another TBG sample with similar twist (Supplementary information). Surprisingly, we did not measure any signal at the scale of the moiré despite the presence of sizeable intervalley scattering at the atomic scale in the top graphene layer. This has to be linked to the internal details of the defects (term $\mathcal{T}$ in Eq.\ref{Eq : LDOS_Theo}). Indeed, tight-binding calculations of Fig.~\ref{fig: TB}b-f reveal that while the $2q-$arcs are universal, there can be order of magnitude variations in the amplitude of the signal depending on the details. While the exact nature of the defect we have observed is unknown, it has been crucial to make the QPIs measurable in our experiment.

In conclusion, our work settles experimentally the band structure of twisted graphene layers of moderately low twist angle where interactions are negligible. We anticipate that QPIs could also be useful to investigate the magic angle regime where strong correlations play an important role.

\bibliographystyle{nature.bst}
\bibliography{Biblio_QPI_bilayer}

\begin{thebibliography}{44}
\expandafter\ifx\csname natexlab\endcsname\relax\def\natexlab#1{#1}\fi
\expandafter\ifx\csname url\endcsname\relax
  \def\url#1{\texttt{#1}}\fi
\expandafter\ifx\csname urlprefix\endcsname\relax\def\urlprefix{URL }\fi

\bibitem[{Bernevig \& Efetov(2024)}]{Bernevig2024}
Bernevig, B.~A. \& Efetov, D.~K.
\newblock {Twisted bilayer graphene’s gallery of phases}.
\newblock \emph{Physics Today} \textbf{77}, 38--44 (2024).
\newblock \urlprefix\url{https://doi.org/10.1063/pt.jvsd.yhyd}.

\bibitem[{Utama \emph{et~al.}(2021)}]{Utama2021}
Utama, M. I.~B. \emph{et~al.}
\newblock Visualization of the flat electronic band in twisted bilayer graphene
  near the magic angle twist.
\newblock \emph{Nature Physics} \textbf{17}, 184--188 (2021).
\newblock \urlprefix\url{https://doi.org/10.1038/s41567-020-0974-x}.

\bibitem[{Lisi \emph{et~al.}(2021)}]{Lisi2021}
Lisi, S. \emph{et~al.}
\newblock Observation of flat bands in twisted bilayer graphene.
\newblock \emph{Nature Physics} \textbf{17}, 189--193 (2021).
\newblock \urlprefix\url{https://doi.org/10.1038/s41567-020-01041-x}.

\bibitem[{Nunn \emph{et~al.}(2023)}]{Nunn2023}
Nunn, J.~E. \emph{et~al.}
\newblock ARPES Signatures of Few-Layer Twistronic Graphenes.
\newblock \emph{Nano Letters} \textbf{23}, 5201--5208 (2023).
\newblock \urlprefix\url{https://doi.org/10.1021/acs.nanolett.3c01173}, pMID:
  37235208.

\bibitem[{Brihuega \emph{et~al.}(2008)}]{Brihuega2008}
Brihuega, I. \emph{et~al.}
\newblock Quasiparticle Chirality in Epitaxial Graphene Probed at the Nanometer
  Scale.
\newblock \emph{Phys. Rev. Lett.} \textbf{101}, 206802 (2008).
\newblock
  \urlprefix\url{http://link.aps.org/doi/10.1103/PhysRevLett.101.206802}.

\bibitem[{Mallet \emph{et~al.}(2012)}]{Mallet2012}
Mallet, P. \emph{et~al.}
\newblock Role of pseudospin in quasiparticle interferences in epitaxial
  graphene probed by high-resolution scanning tunneling microscopy.
\newblock \emph{Phys. Rev. B} \textbf{86}, 045444 (2012).
\newblock \urlprefix\url{http://link.aps.org/doi/10.1103/PhysRevB.86.045444}.

\bibitem[{Dutreix \& Katsnelson(2016)}]{Dutreix2016}
Dutreix, C. \& Katsnelson, M.~I.
\newblock Friedel oscillations at the surfaces of rhombohedral $N$-layer
  graphene.
\newblock \emph{Phys. Rev. B} \textbf{93}, 035413 (2016).
\newblock \urlprefix\url{http://dx.doi.org/10.1103/PhysRevB.93.035413}.

\bibitem[{Dutreix \& \textit{et al}(2019)}]{Dutreix2019}
Dutreix, C. \& \textit{et al}.
\newblock Measuring the Berry phase of graphene from wavefront dislocations in
  Friedel oscillations.
\newblock \emph{Nature} \textbf{574}, 219–222 (2019).

\bibitem[{Zhang \emph{et~al.}(2020)Zhang, Su \& He}]{Zhang2020}
Zhang, Y., Su, Y. \& He, L.
\newblock Local Berry Phase Signatures of Bilayer Graphene in Intervalley
  Quantum Interference.
\newblock \emph{Phys. Rev. Lett.} \textbf{125}, 116804 (2020).
\newblock
  \urlprefix\url{https://link.aps.org/doi/10.1103/PhysRevLett.125.116804}.

\bibitem[{Nuckolls \emph{et~al.}(2023)}]{Nuckolls2023}
Nuckolls, K.~P. \emph{et~al.}
\newblock Quantum textures of the many-body wavefunctions in magic-angle
  graphene.
\newblock \emph{Nature} \textbf{620}, 525–532 (2023).

\bibitem[{Guan \& \textit{et al}(2024)}]{Guan2024}
Guan, U. \& \textit{et al}.
\newblock Observation of Kekulé vortices around hydrogen adatoms in graphene.
\newblock \emph{Nature Commun} \textbf{15}, 2927 (2024).

\bibitem[{Holbrook \emph{et~al.}(2024)}]{holbrook2024}
Holbrook, M. \emph{et~al.}
\newblock Real-Space Imaging of the Band Topology of Transition Metal
  Dichalcogenides (2024).
\newblock \urlprefix\url{https://arxiv.org/abs/2412.02813}.

\bibitem[{Călugăru \emph{et~al.}(2025)}]{calugaru2025}
Călugăru, D. \emph{et~al.}
\newblock Probing the Quantized Berry Phases in 1H-NbSe$_2$ Using Scanning
  Tunneling Microscopy (2025).
\newblock \urlprefix\url{https://arxiv.org/abs/2501.09063}.

\bibitem[{de~Gail \emph{et~al.}(2011)de~Gail, Goerbig, Guinea, Montambaux \&
  Castro~Neto}]{deGail2011}
de~Gail, R., Goerbig, M.~O., Guinea, F., Montambaux, G. \& Castro~Neto, A.~H.
\newblock Topologically protected zero modes in twisted bilayer graphene.
\newblock \emph{Phys. Rev. B} \textbf{84}, 045436 (2011).
\newblock \urlprefix\url{https://link.aps.org/doi/10.1103/PhysRevB.84.045436}.

\bibitem[{Lopes~dos Santos \emph{et~al.}(2007)Lopes~dos Santos, Peres \&
  Castro~Neto}]{LopesdosSantos2007}
Lopes~dos Santos, J. M.~B., Peres, N. M.~R. \& Castro~Neto, A.~H.
\newblock Graphene Bilayer with a Twist: Electronic Structure.
\newblock \emph{Phys. Rev. Lett.} \textbf{99}, 256802 (2007).
\newblock
  \urlprefix\url{https://link.aps.org/doi/10.1103/PhysRevLett.99.256802}.

\bibitem[{Bistritzer \& MacDonald(2011)}]{Bistritzer2011}
Bistritzer, R. \& MacDonald, A.~H.
\newblock Moiré bands in twisted double-layer graphene.
\newblock \emph{PNAS} \textbf{108}, 12233 12237 (2011).
\newblock
  \urlprefix\url{https://www.pnas.org/doi/full/10.1073/pnas.1108174108}.

\bibitem[{Zou \emph{et~al.}(2018)Zou, Po, Vishwanath \& Senthil}]{Zou2018}
Zou, L., Po, H.~C., Vishwanath, A. \& Senthil, T.
\newblock Band structure of twisted bilayer graphene: Emergent symmetries,
  commensurate approximants, and Wannier obstructions.
\newblock \emph{Phys. Rev. B} \textbf{98}, 085435 (2018).
\newblock \urlprefix\url{https://link.aps.org/doi/10.1103/PhysRevB.98.085435}.

\bibitem[{Song \emph{et~al.}(2019)}]{Song2019}
Song, Z. \emph{et~al.}
\newblock All Magic Angles in Twisted Bilayer Graphene are Topological.
\newblock \emph{Phys. Rev. Lett.} \textbf{123}, 036401 (2019).
\newblock
  \urlprefix\url{https://link.aps.org/doi/10.1103/PhysRevLett.123.036401}.

\bibitem[{Cao \emph{et~al.}(2019{\natexlab{a}})}]{Cao2019b}
Cao, Y. \emph{et~al.}
\newblock Correlated insulator behaviour at half-filling in magic-angle
  graphene superlattices.
\newblock \emph{Nature} \textbf{556}, 80 (2019{\natexlab{a}}).

\bibitem[{Lu \emph{et~al.}(2019)Lu, Stepanov, Yang \& \textit{et al.}}]{Lu2019}
Lu, X., Stepanov, P., Yang, W. \& \textit{et al.}
\newblock Superconductors, orbital magnets and correlated states in magic-angle
  bilayer graphene.
\newblock \emph{Nature} \textbf{574}, 653 (2019).

\bibitem[{Cao \emph{et~al.}(2019{\natexlab{b}})}]{Cao2019}
Cao, Y. \emph{et~al.}
\newblock Magic-angle graphene superlattices: a new platform for unconventional
  superconductivity.
\newblock \emph{Nature} \textbf{55}, 43--50 (2019{\natexlab{b}}).

\bibitem[{Stepanov \emph{et~al.}(2020)}]{Stepanov2020}
Stepanov, P. \emph{et~al.}
\newblock Decoupling superconductivity and correlated insulators in twisted
  bilayer graphene.
\newblock \emph{Nature} \textbf{583}, 375 (2020).

\bibitem[{Cao \emph{et~al.}(2020)}]{Cao2020}
Cao, Y. \emph{et~al.}
\newblock Strange Metal in Magic-Angle Graphene with near Planckian
  Dissipation.
\newblock \emph{Phys. Rev. Lett.} \textbf{124}, 076801 (2020).

\bibitem[{Lu \emph{et~al.}(2022)Lu, Das, Di~Battista \& \textit{et
  al.}}]{Jaoui2022}
Lu, X., Das, I., Di~Battista, G. \& \textit{et al.}
\newblock Quantum critical behaviour in magic-angle twisted bilayer graphene.
\newblock \emph{Nat. Phys.} \textbf{18}, 633 (2022).

\bibitem[{Serlin \emph{et~al.}(2020)}]{Serlin2020}
Serlin, M. \emph{et~al.}
\newblock Intrinsic quantized anomalous Hall effect in a moir{\'e}
  heterostructure.
\newblock \emph{Science} \textbf{367}, 900--903 (2020).
\newblock \urlprefix\url{https://science.sciencemag.org/content/367/6480/900}.

\bibitem[{Nuckolls \emph{et~al.}(2020)}]{Nuckolls2020}
Nuckolls, K.~P. \emph{et~al.}
\newblock Strongly Correlated Chern Insulators in Magic-Angle Twisted Bilayer
  Graphene.
\newblock \emph{Nature} \textbf{588}, 610 (2020).

\bibitem[{Wu \emph{et~al.}(2021)Wu, Zhang, Watanabe, Taniguchi \& E.}]{Wu2021}
Wu, S., Zhang, Z., Watanabe, K., Taniguchi, T. \& E., A.
\newblock Chern insulators, van Hove singularities and topological flat bands
  in magic-angle twisted bilayer graphene.
\newblock \emph{Nat. Mater} \textbf{20}, 488 (2021).

\bibitem[{Xie \& \textit{et al.}(2021)}]{Xie2021}
Xie, Y. \& \textit{et al.}
\newblock Fractional Chern insulators in magic-angle twisted bilayer graphene.
\newblock \emph{Nature} \textbf{600}, 439 (2021).

\bibitem[{Yu \emph{et~al.}(2022)Yu, Foutty, Han \& \textit{et al.}}]{Yu2022}
Yu, J., Foutty, B.~A., Han, Z. \& \textit{et al.}
\newblock Correlated Hofstadter spectrum and flavour phase diagram in
  magic-angle twisted bilayer graphene.e.
\newblock \emph{Nat. Phys} \textbf{18}, 835 (2022).

\bibitem[{Tian \emph{et~al.}(2023)Tian, Gao, Zhang \& \textit{et
  al.}}]{Tian2023}
Tian, H., Gao, X., Zhang, Y. \& \textit{et al.}
\newblock Evidence for Dirac flat band superconductivity enabled by quantum
  geometry.
\newblock \emph{Nature} \textbf{614}, 440--444 (2023).

\bibitem[{Tanaka \emph{et~al.}(2025)Tanaka, Wang, Dinh \& \textit{et
  al.}}]{Tanaka2025}
Tanaka, M., Wang, J., Dinh, T. \& \textit{et al.}
\newblock Superfluid stiffness of magic-angle twisted bilayer graphene.
\newblock \emph{Nature} \textbf{638}, 99--105 (2025).

\bibitem[{Phong \& Mele(2020)}]{Mele2020}
Phong, V.~o.~T. \& Mele, E.~J.
\newblock Obstruction and Interference in Low-Energy Models for Twisted Bilayer
  Graphene.
\newblock \emph{Phys. Rev. Lett.} \textbf{125}, 176404 (2020).
\newblock
  \urlprefix\url{https://link.aps.org/doi/10.1103/PhysRevLett.125.176404}.

\bibitem[{Kumar \emph{et~al.}(2016)}]{Kumar2016}
Kumar, B. \emph{et~al.}
\newblock Growth Protocols and Characterization of Epitaxial Graphene on
  {{SiC}} Elaborated in a Graphite Enclosure.
\newblock \emph{Phys. E Low-Dimens. Syst. Nanostructures} \textbf{75}, 7--14
  (2016).

\bibitem[{Simon \emph{et~al.}(2007)Simon, Vonau \& Aubel}]{Simon2007}
Simon, L., Vonau, F. \& Aubel, D.
\newblock A phenomenological approach of joint density of states for the
  determination of band structure in the case of a semi-metal studied by
  FT-STS.
\newblock \emph{Journal of Physics: Condensed Matter} \textbf{19}, 355009
  (2007).
\newblock \urlprefix\url{https://dx.doi.org/10.1088/0953-8984/19/35/355009}.

\bibitem[{Kang \& Vafek(2018)}]{Kang2018}
Kang, J. \& Vafek, O.
\newblock Symmetry, Maximally Localized Wannier States, and a Low-Energy Model
  for Twisted Bilayer Graphene Narrow Bands.
\newblock \emph{Phys. Rev. X} \textbf{8}, 031088 (2018).
\newblock \urlprefix\url{https://link.aps.org/doi/10.1103/PhysRevX.8.031088}.

\bibitem[{Koshino \emph{et~al.}(2018)}]{Koshino2018}
Koshino, M. \emph{et~al.}
\newblock Maximally Localized Wannier Orbitals and the Extended Hubbard Model
  for Twisted Bilayer Graphene.
\newblock \emph{Phys. Rev. X} \textbf{8}, 031087 (2018).
\newblock \urlprefix\url{https://link.aps.org/doi/10.1103/PhysRevX.8.031087}.

\bibitem[{Dutreix \emph{et~al.}(2021)}]{Dutreix2021}
Dutreix, C. \emph{et~al.}
\newblock Measuring graphene{\textquoteright}s {Berry} phase at $B=0${~T}.
\newblock \emph{Comptes Rendus. Physique} \textbf{22}, 133--143 (2021).

\bibitem[{Trambly~de Laissardière \emph{et~al.}(2010)Trambly~de Laissardière,
  Mayou \& Magaud}]{Trambly2010}
Trambly~de Laissardière, G., Mayou, D. \& Magaud, L.
\newblock Localization of Dirac Electrons in Rotated Graphene Bilayers.
\newblock \emph{Nano Letters} \textbf{10}, 804--808 (2010).
\newblock \urlprefix\url{https://doi.org/10.1021/nl902948m}.

\bibitem[{Huder \emph{et~al.}(2018)}]{Huder2018}
Huder, L. \emph{et~al.}
\newblock Electronic spectrum of twisted graphene layers under heterostrain.
\newblock \emph{Physical Review Letters} \textbf{120}, 156405 (2018).

\bibitem[{Fukui \emph{et~al.}(2005)Fukui, Hatsugai \& Suzuki}]{Fukui2005}
Fukui, T., Hatsugai, Y. \& Suzuki, H.
\newblock Chern Numbers in Discretized Brillouin Zone: Efficient Method of
  Computing (Spin) Hall Conductances.
\newblock \emph{Journal of the Physical Society of Japan} \textbf{74},
  1674--1677 (2005).
\newblock \urlprefix\url{https://doi.org/10.1143/JPSJ.74.1674}.

\bibitem[{Mesple \emph{et~al.}(2021)}]{Mesple2021}
Mesple, F. \emph{et~al.}
\newblock Heterostrain Determines Flat Bands in Magic-Angle Twisted Graphene
  Layers.
\newblock \emph{Phys. Rev. Lett.} \textbf{127}, 126405 (2021).

\bibitem[{Herzog-Arbeitman \emph{et~al.}(2024)}]{Herzog2024}
Herzog-Arbeitman, J. \emph{et~al.}
\newblock Heavy Fermions as an Efficient Representation of Atomistic Strain and
  Relaxation in Twisted Bilayer Graphene (2024).
\newblock \urlprefix\url{https://arxiv.org/abs/2405.13880}.

\bibitem[{Bi \emph{et~al.}(2019)Bi, Yuan \& Fu}]{Bi2019}
Bi, Z., Yuan, N. F.~Q. \& Fu, L.
\newblock Designing flat bands by strain.
\newblock \emph{Phys. Rev. B} \textbf{100}, 035448 (2019).

\bibitem[{Lopes~dos Santos \emph{et~al.}(2012)Lopes~dos Santos, Peres \&
  Castro~Neto}]{LopesdosSantos2012}
Lopes~dos Santos, J. M.~B., Peres, N. M.~R. \& Castro~Neto, A.~H.
\newblock Continuum model of the twisted graphene bilayer.
\newblock \emph{Phys. Rev. B} \textbf{86}, 155449 (2012).
\newblock \urlprefix\url{https://link.aps.org/doi/10.1103/PhysRevB.86.155449}.

\end{thebibliography}

\section*{Methods}
\subsection*{STM measurements}
The measurements were acquired in a custom-built UHV STM in a cryogenic environment at 8.4 K and 10$^{-10}$ mbar. Thermal broadening results in a 2.6meV broadening that does not affect QPI interpretation. The STM tip is a wire-cut Pt/Ir, the apex is cleaned with field emission and prepared on Ag(111). Density of states maps are taken using Lock-In phase sensitive detection with a 10 mV AC voltage modulation.

\subsection*{Sample fabrication}
Multilayer graphene was grown on the C-face of 6H-SiC, a custom built RF-furnace.\cite{Kumar2016} During the first step of the process, SiC was hydrogen etched by exposing it to to a 98\%Ar 2\%H2 dynamic flow, at a temperature of 1600$^{\circ}$C for 30mn (ramps are 1h long). Then the graphene was grown on the etched C-face SiC at 1600$^{\circ}$C for 30mn, with slow ramp rates (3 \& 5 hours for the up and down ramps respectively). The sample was then further annealed in the UHV-STM chamber for degassing.


\section*{Acknowledgements}
V.T.R and G. Trambly de Laissardière acknowledge the support from the ANR Flatmoi project (ANR-21-CE30-0029). TB calculations have been performed at the Centre de Calcul (CDC), CY Cergy Paris Universit\'e, and at TGCC-GENCI (Project AD010910784). C.D. acknowledges support from the project TopoMat (ANR-23-CE30-0029) funded by the French
Research National Agency. 

\section*{Author contributions}
F.M, P.M J-Y.V and V.T.R conceived the experiments. F.M, G.L, P.M and J-Y.V performed the experiments. F.M, V.T.R, P.M and J-Y.V analysed experimental data. G.TdL performed tight-binding calculations and C.D conducted T-matrix analysis of conductance maps. V.T.R wrote the manuscript with the input of all authors and coordinated the collaboration.


	
\end{document}